\documentclass[a4paper,11pt]{article}

\usepackage{jcappub, bm, color} 
\usepackage{amssymb,amsfonts,slashed,amsthm,amsmath,graphicx, soul}
\usepackage[caption=false]{subfig}
\usepackage{epsfig}

\begin{document}


\newcommand{\vev}[1]{ \left\langle {#1} \right\rangle }
\newcommand{\bra}[1]{ \langle {#1} | }
\newcommand{\ket}[1]{ | {#1} \rangle }
\newcommand{\eV}{ \ {\rm eV} }
\newcommand{\KeV}{ \ {\rm keV} }
\newcommand{\MeV}{\  {\rm MeV} }
\newcommand{\GeV}{\  {\rm GeV} }
\newcommand{\TeV}{\  {\rm TeV} }
\newcommand{\1}{\mbox{1}\hspace{-0.25em}\mbox{l}}
\newcommand{\Red}[1]{{\color{red} {#1}}}

\newcommand{\lmk}{\left(}  
\newcommand{\rmk}{\right)}
\newcommand{\lkk}{\left[}  
\newcommand{\rkk}{\right]}
\newcommand{\lhk}{\left \{ }  
\newcommand{\rhk}{\right \} }
\newcommand{\del}{\partial}  
\newcommand{\la}{\left\langle} 
\newcommand{\ra}{\right\rangle}
\newcommand{\half}{\frac{1}{2}}

\newcommand{\bea}{\begin{array}}
\newcommand{\eea}{\end{array}}
\newcommand{\beq}{\begin{eqnarray}}
\newcommand{\eeq}{\end{eqnarray}}
\newcommand{\eq}[1]{Eq.~(\ref{#1})}

\newcommand{\dd}{\mathrm{d}}
\newcommand{\Mpl}{M_{p}}
\newcommand{\mg}{m_{3/2}}
\newcommand{\abs}[1]{\left\vert {#1} \right\vert}
\newcommand{\mphi}{m_{\phi}}
\newcommand{\Hz}{\ {\rm Hz}}
\newcommand{\for}{\quad \text{for }}
\newcommand{\Min}{\text{Min}}
\newcommand{\Max}{\text{Max}}
\newcommand{\Kahler}{K\"{a}hler }
\newcommand{\cphi}{\varphi}
\newcommand{\Tr}{\text{Tr}}
\newcommand{\diag}{{\rm diag}}

\newcommand{\ubar}{u^c}
\newcommand{\dbar}{d^c}
\newcommand{\ebar}{e^c}
\newcommand{\nubar}{\nu^c}
\newcommand{\Ndw}{N_{\rm DW}}
\newcommand{\Fpq}{F_{\rm PQ}}
\newcommand{\fpq}{v_{\rm PQ}}
\newcommand{\Br}{{\rm Br}}
\newcommand{\Lag}{\mathcal{L}}
\newcommand{\Lqcd}{\Lambda_{\rm QCD}}

\newcommand{\im}{{\rm Im} }
\newcommand{\re}{{\rm Re} }

\def\lrf#1#2{ \left(\frac{#1}{#2}\right)}
\def\lrfp#1#2#3{ \left(\frac{#1}{#2} \right)^{#3}}
\def\lrp#1#2{\left( #1 \right)^{#2}}
\def\REF#1{Ref.~\cite{#1}}
\def\SEC#1{Sec.~\ref{#1}}
\def\FIG#1{Fig.~\ref{#1}}
\def\EQ#1{Eq.~(\ref{#1})}
\def\EQS#1{Eqs.~(\ref{#1})}
\def\TEV#1{10^{#1}{\rm\,TeV}}
\def\GEV{\rm\,GeV}
\def\MEV{\rm\,MeV}
\def\KEV{\rm\,keV}
\def\EV{\rm\,eV}
\def\blue#1{\textcolor{blue}{#1}}
\def\red#1{\textcolor{blue}{#1}}

\def\({\left(}
\def\){\right)}
\def\a{\alpha}
\def\b{\beta}
\def\c{\varepsilon}
\def\d{\delta}
\def\e{\epsilon}
\def\f{\phi}
\def\g{\gamma}
\def\h{\theta}
\def\k{\kappa}
\def\m{\mu}
\def\n{\nu}
\def\p{\psi}
\def\q{\partial}
\def\r{\rho}
\def\s{\sigma}
\def\t{\tau}
\def\u{\upsilon}
\def\w{\omega}
\def\x{\xi}
\def\y{\eta}
\def\z{\zeta}
\def\D{\Delta}
\def\G{\Gamma}
\def\H{\Theta}
\def\F{\Phi}
\def\P{\Psi}
\def\S{\Sigma}
\def\me{\mathrm e}
\def\ol{\overline}
\def\tl{\tilde}
\def\*{\dagger}
\newcommand{\OR}{~{\rm or}~}
\newcommand{\AND}{~{\rm and}~}
\newcommand{\WITH}{ ~{\rm with}~ }
\def\O{\mathcal{O}}
\def\->{\rightarrow}
\def\<-{\leftarrow}
\newcommand{\laq}[1]{\label{#1}}


\begin{flushright}
TU-1106, IPMU20-0081
\end{flushright}

\title{
What if ALP dark matter for the XENON1T excess is the inflaton
}

\author{
Fuminobu Takahashi$^{1,2}$, Masaki Yamada$^{1,3}$, 
and Wen Yin$^4$
\\*[20pt]
$^1${\it \normalsize 
Department of Physics, Tohoku University, 
Sendai, Miyagi 980-8578, Japan} \\*[5pt]
$^2${\it \normalsize
Kavli IPMU (WPI), UTIAS, 
The University of Tokyo, 
Kashiwa, Chiba 277-8583, Japan} \\*[5pt]
$^3${\it \normalsize
Frontier Research Institute for Interdisciplinary Sciences, Tohoku University, \\
Sendai, Miyagi 980-8578, Japan
} \\*[5pt]
$^4${\it \normalsize
Department of Physics, Faculty of Science, The University of Tokyo,  \\ 
Bunkyo-ku, Tokyo 113-0033, Japan
}
\vspace{0.5cm}
}

\emailAdd{fumi@tohoku.ac.jp}
\emailAdd{m.yamada@tohoku.ac.jp}
\emailAdd{yinwen@hep-th.phys.s.u-tokyo.ac.jp,  yinwen@tuhep.phys.tohoku.ac.jp}

\abstract{
The recent XENON1T excess in the electron recoil data can be explained by anomaly-free axion-like particle (ALP) dark matter with mass $m_\phi = 2.3 \pm 0.2$\,keV and the decay constant $f_\phi/q_e \simeq 2 \times 10^{10}  \sqrt{\Omega_\phi/\Omega_{\rm DM}}{\rm \, GeV}$. Intriguingly, the suggested mass and decay constant are consistent with the relation, $f_\phi \sim 10^3 \sqrt{m_\phi M_p}$, predicted in a scenario where the ALP plays the role of the inflaton. This raises a possibility that the ALP dark matter responsible for the XENON1T excess also drove inflation in the very early universe. We study implications of the XENON1T excess for the ALP inflation and thermal history of the universe after inflation.
}

\maketitle

\section{Introduction
\label{sec:introduction}}
Recently, an excess was found in the electron recoil data collected by the XENON1T experiment~\cite{Aprile:2020tmw}, which stimulated active discussion on the possible explanation for the excess.
While the tritium is not excluded as a mundane explanation\footnote{It was pointed out
in Ref.~\cite{Szydagis:2020isq} that ${}^{37}$Ar can also explain the excess.}, it is worth studying the interpretation in terms of new physics beyond the standard model (SM). 

One of such is the axion-like particle (ALP) dark matter (DM)~\cite{Takahashi:2020bpq, Bloch:2020uzh,Athron:2020maw} (see Refs.~\cite{Jaeckel:2010ni, Ringwald:2012hr, Arias:2012az, Graham:2015ouw, Marsh:2015xka, Irastorza:2018dyq, DiLuzio:2020wdo} for reviews), which generates a mono-energetic peak through absorption by electron.
The XENON1T excess favors the ALP mass and coupling to electron in the range of
\begin{align}
\label{mass}
    m_\phi &= 2.3 \pm 0.2 {\rm\,keV}\,\,(68\%{\rm\,C.L.}),\\
    \label{gphie}
    g_{\phi e} & \simeq 3 \times 10^{-14} \sqrt{\frac{\Omega_{\rm DM}}{\Omega_{\phi}}},
\end{align}
with a $3.0\, \sigma$ global ($4.0\, \sigma$ local) significance over background~\cite{Aprile:2020tmw}, where $\Omega_{\rm DM}$ and $\Omega_\phi$ are the density parameters of DM and ALP $\phi$, respectively.\footnote{Dark photon DM~\cite{Alonso-Alvarez:2020cdv,Nakayama:2020ikz,An:2020bxd} can also explain the  excess with 
the same global significance.
}
Let us define the fraction of ALP DM by $r \equiv \Omega_\phi/\Omega_{\rm DM}$.
One can express the coupling to electrons in terms of the decay constant
as
\begin{align}
    f_\phi &\simeq 2 \times 10^{10} q_e \sqrt{r} {\rm \,GeV},
    \label{fphi}
\end{align}
where $q_e$ is the Peccei-Quinn (PQ) charge of the electron.
Since the ALP generically has an anomalous coupling to photons, its decay produces X-ray line signal at an energy
equal to half the ALP mass. To satisfy the X-ray observations for the above mass and decay constant, however,
the ALP coupling to photons
must be extremely suppressed than naively expected~\cite{Pospelov:2008jk,Arias:2012az,Nakayama:2014cza,Takahashi:2020bpq}.
One way to suppress the photon coupling is to consider an anomaly-free ALP which has no QED anomaly~\cite{Nakayama:2014cza,Takahashi:2020bpq}\footnote{\label{ft1}The ALP should not have QCD anomaly, either, to avoid
a mixing with $\pi^0$~\cite{Takahashi:2020bpq}. In particular, the kinetic or mass mixing between the ALP and the QCD axion should be suppressed.}.
Then, the dominant coupling to photons arises from the electron threshold of the form, 
\beq 
\label{photon} \frac{\alpha }{48 \pi (f_\phi/q_e) } \frac{\partial^2 \phi}{m_e^2} F_{\mu \nu} \tilde{F}^{\mu \nu},
\eeq 
which predicts an X-ray line signal with definite strength~\cite{Nakayama:2014cza,Takahashi:2020bpq} (see also Ref.~\cite{Pospelov:2008jk}).
The predicted X-ray line may be searched for by future X-ray observatories~\cite{Nandra:2013jka,LynxTeam:2018usc,Zhang:2016ach,DeAngelis:2017gra}.

The  coupling to electron (\ref{gphie}) satisfies the stellar cooling bound, and actually it is close to the value
 hinted by the stellar cooling anomalies of white dwarfs (WD)~\cite{Bertolami:2014wua,Corsico:2016okh} and red giants (RG)~\cite{Viaux:2013lha}\footnote{See, however, Ref.~\cite{Capozzi:2020cbu} where 
 they placed a tight upper bound on the coupling to electrons using the Gaia data.}, and the agreement becomes even better if the ALP is a part of the total DM~\cite{Takahashi:2020bpq}. Furthermore, the right abundance of ALP DM can be  generated by the misalignment mechanism, or by thermal scatterings if 
 the ALP constitutes a fraction of DM~\cite{Arias:2012az,Nakayama:2014cza,Takahashi:2020bpq}.
 In this paper we will point out another coincidence of the ALP parameters in a different context.

There is a strong evidence~\cite{Akrami:2018odb} for inflation at an early stage of the universe~\cite{Guth:1980zm,Starobinsky:1980te,Sato:1980yn}. The slow-roll inflation paradigm~\cite{Linde:1981mu,Albrecht:1982wi} crucially depends on a sufficiently flat inflaton potential, and it has been one of the central issues in the inflation model building how to protect the flat potential from radiative corrections since the early discussion on the GUT higgs inflation in the 1980s.
One of the plausible candidates for the inflaton is an axion whose potential is naturally protected by the shift symmetry.

There are a wide variety of inflation models based on the axion. Natural inflation~\cite{Freese:1990rb} is a simple and therefore attractive model in which the inflaton potential consists of a single cosine function. However, the decay constant is required to be super-Planckian for the slow-roll inflation, and its predicted scalar spectral index as well as the tensor-to-scalar ratio are disfavored by the current CMB observations~\cite{Akrami:2018odb}. There are many extensions of the natural inflation
(see e.g. Refs.~\cite{ArkaniHamed:2003wu,Czerny:2014wza,Czerny:2014qqa,Croon:2014dma,Higaki:2015kta,Daido:2017wwb,Daido:2017tbr,Takahashi:2019qmh}), and we focus on the simplest class of the so-called multi-natural inflation~\cite{Czerny:2014wza}; the inflaton potential consists of two cosine terms which conspire to make a flat plateau around the potential maximum.
This potential allows slow-roll inflation from very high energy scales to very low energy scales depending on the value of the decay constant. Such axion inflation model may be realized in the axion landscape~\cite{Higaki:2014mwa,Higaki:2014pja}
where many axions have various shift symmetry breaking terms and mixing.

In the simplest class of the multi-natural inflation,
the inflaton potential often has an upside-down symmetry, where
the curvatures of the potential maximum and minimum are of equal magnitude but opposite sign (see Fig.~\ref{fig:1}). 
Then, the inflaton mass at the minimum is naturally suppressed once we require the slow-roll conditions
to be satisfied around the potential maximum.
Thanks to the light mass, the inflaton can be stable on the cosmological time scale 
and contribute to DM.
In Refs.~\cite{Daido:2017wwb, Daido:2017tbr}, Daido and two of the present authors (FT and WY) investigated this possibility. 
A consistency relation between the ALP mass and decay constant was derived solely based on the CMB normalization of the density perturbation and the observed scalar spectral index, and it reads
\begin{align}
\label{ALPinfrel}
    f_\phi \sim 10^3 \sqrt{m_\phi M_{p}},
\end{align}
where $M_p \simeq 2.4 \times 10^{18}$\,GeV is the reduced Planck mass (see also Ref.~\cite{Czerny:2014xja}). This is a rather robust relation. Based on this set-up, it led to the so-called ALP miracle scenario in which the ALP DM and inflation are explained in a unified manner, and the suggested parameter space is within the reach of future solar axion experiments such as IAXO~\cite{Armengaud:2014gea,Armengaud:2019uso}. Intriguingly, the parameters (\ref{mass}) and (\ref{gphie}) suggested by the XENON1T excess satisfy the above relation. This coincidence implies that the ALP DM, which is responsible for the XENON1T excess, might have played the role of the inflaton in the very early universe. 

In this paper we study cosmological implications of the coincidence between the XENON1T excess and the predicted relation between the ALP mass and decay constant in a context of the ALP inflation.
In the next section, we explain the ALP inflation model 
and the origin of \eq{ALPinfrel}. 
We also discuss the reheating of the ALP inflation 
and derive the condition that the coherent oscillation of the ALP inflaton is diffused into the thermal plasma. 
The ALPs are, however, produced from the scattering in the thermal plasma during the diffusion process. 
These ALPs are the potential source of the XENON1T excess. 
In Sec.~\ref{sec:entropy}, we discuss an example of the scenario for a mild entropy production that is required to  dilute the thermally produced ALP so that the abundance of the thermally produced ALP does not exceeds a present upper bound. 
In Sec.~\ref{sec:discussion}, we discuss the possibilities with multiple ALPs. 
In particular, we comment on anthropic explanation of the ALP mass in the context of string landscape. 
Section~\ref{sec:conclusion} is devoted to conclusion.

\section{ALP inflation and XENON1T excess
\label{sec:inflation}}
\subsection{ALP inflation}

\begin{figure}[!t]
  \begin{center}
   \includegraphics[width=105mm]{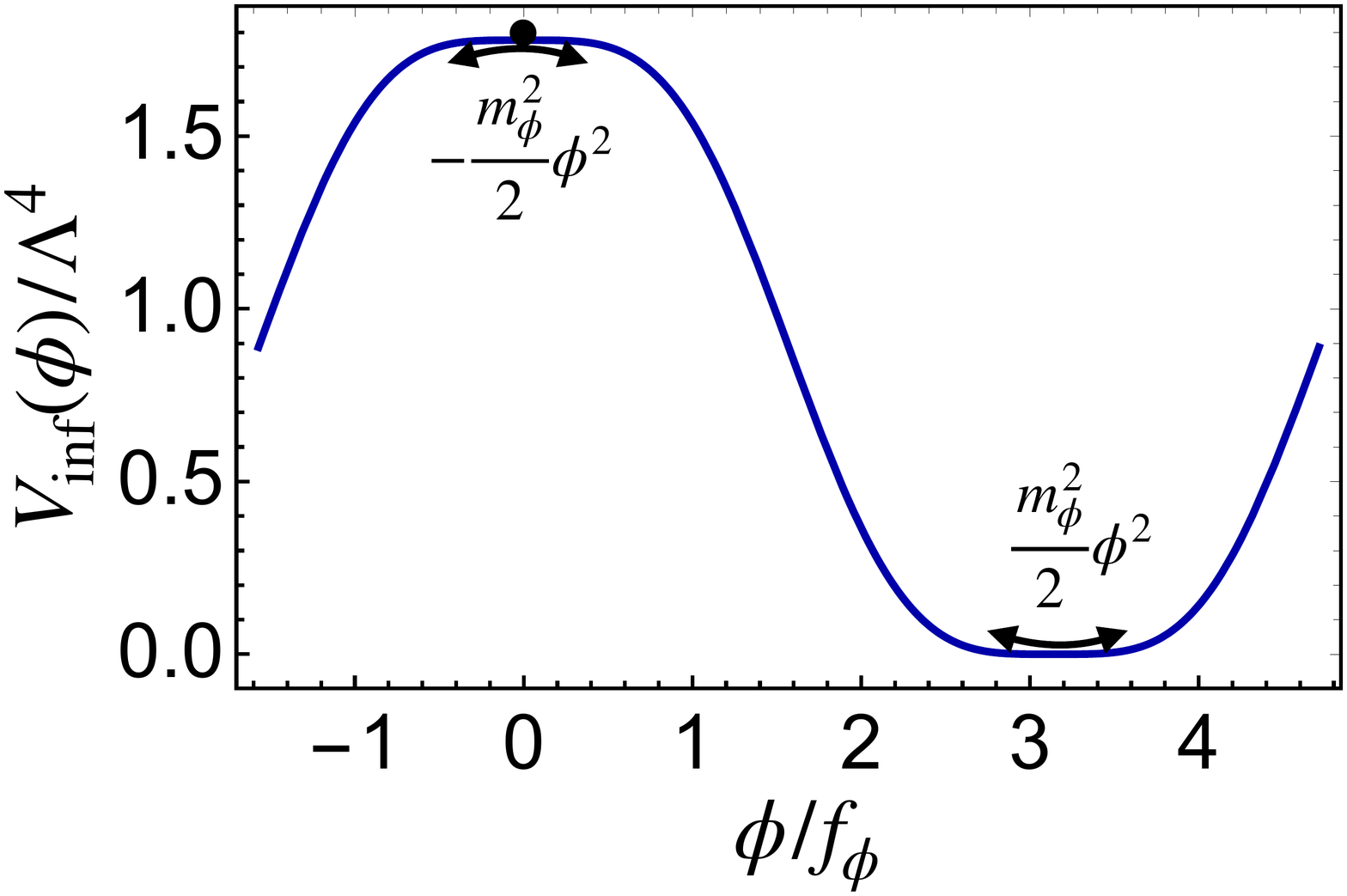}
\end{center}
\caption{
The ALP potential for $n=3$ with $\theta \approx 0$ and $\kappa \approx 1$.
The curvatures of the potential at the maximum and minimum are equal to each other but with an opposite sign. 
The slow-roll inflation takes place around the hilltop of the potential while the ALP excitation around the
minimum contributes to DM. }
\label{fig:1}
\end{figure}

We assume that the ALP potential consists of 
two sinusoidal functions such as~\cite{Czerny:2014wza}
\begin{align}
\label{DIV} 
V_{\rm inf}(\phi) = 
\Lambda^4\left(\cos\left({\frac{\phi}{ f_\phi }} + \theta \right)-{\frac{\kappa}{n^2}}\cos\left({\frac{n\phi}{f_\phi} }\right)
\right)
+{\rm const.},
\end{align}
where $n (>1)$ is a rational number, $\kappa$ is a numerical coefficient, and $\theta$ is a relative phase. 
We have added the last constant term so that the cosmological constant is vanishingly small in the present vacuum. 
If $n$ is an odd integer, the potential has an upside-down symmetry (See Fig. \ref{fig:1}). In particular, the mass squared of the ALP at the potential minimum is equal to the curvature at the potential maximum but with an opposite sign. In the following we assume the slow-roll towards $\phi>0$ without loss of generality.

For the moment we set $\theta = 0$ and $\kappa = 1$ so that the curvature vanishes, $V''(0)=0$,  at the origin. 
In this limit, the inflaton dynamics is reduced to
the standard hilltop quartic inflation, and the inflaton evolution
can be solved analytically. 
Then, the potential near the origin is  approximated by
\beq
\label{Vinf}
V_{\rm inf}(\phi) \simeq   V_0-  \lambda {\phi^4 } 
+ \cdots,
\eeq
where we have defined
 \begin{align}
 \label{V0}
 V_0 &\equiv  V_{\rm inf}(0) - V_{\rm inf}(\pi f_\phi) 
 = 2{\frac{n^2- 1}{n^2}} \Lambda^4 \\
 \lambda & \equiv {\frac{n^2- 1}{4!}}\({\frac{\Lambda}{ f_\phi}}\)^4.
  \label{lambda}
\end{align}
Here the potential contains no linear or quadratic terms, but the inflaton dynamics is not significantly
modified as long as these terms are sufficiently small. In fact, we as shall see shortly, while the hilltop quartic inflation
can last sufficiently long to solve the theoretical  problems of the standard big bang cosmology, 
its prediction of the spectral index is too small to explain the observed value. Then,
we will introduce linear and quadratic terms as small corrections, which give a better fit to
the observed spectral index.\footnote{A cubic term is also induced in our set-up, but it does not modify the inflaton dynamics
significantly.} 

The amplitude of curvature perturbations is calculated from 
\beq
\laq{pln}
P_{\mathcal{R}} = \(\frac{H_{\rm inf}^2}{2\pi \dot{\phi}_*}\)^2 \simeq 
\frac{V_{\rm inf}(\phi_*)^3}{12 \pi^2 V_{\rm inf}'(\phi_*)^2 M_p^6},
\eeq 
where the subscript $*$ implies that the variable is evaluated at the horizon exit of
the pivot scale $k_*=0.05 {\rm Mpc}^{-1}$.
The observed amplitude of the curvature perturbation is given by~\cite{Akrami:2018odb}
\begin{align}
\label{pn}
P_{\mathcal{R}} &\simeq 2.1 \times 10^{-9}. 
\end{align}
This fixes the quartic coupling as
\beq
\label{pninni}
\lambda \simeq  7.5 \times 10^{-14} \({\frac{N_*}{ 50}}\)^{-3}, 
\eeq
where $N_*$ is the e-folding number at the horizon exit of the pivot scale.

The scalar spectral index $n_s$ is given by
\begin{equation}
\label{eq:nsform}
n_s \simeq  1- 6 \varepsilon + 2\eta,
\end{equation}
 where  the slow-roll parameters are defined as
\begin{align}
\varepsilon(\phi) &\equiv \frac{M_p^2}{2} \left(\frac{V_{\rm inf}'}{V_{\rm inf}}\right)^2,\\
\eta({\phi})  &\equiv M_{p}^2 \frac{V_{\rm inf}''}{V_{\rm inf}}.
\end{align}
Unless the inflation scale is very high, the contribution to $n_s$ is known to be dominated by $\eta$.
In the hilltop quartic inflation, the inflaton dynamics can 
be easily solved analytically and one obtains $\eta(\phi_*) \simeq - 3/2 N_*$, and thus,
\beq
 n_s(\phi_*) \simeq 1-{\frac{3}{N_*}}.
\eeq 
We are interested in the case of the Hubble parameter during inflation of order keV, which results in $N_* = 32 \, \text{-} \, 35$. Then, the predicted value of the spectral index in the quartic hilltop inflation is too small to explain the observed value, $n_s =0.9649 \pm 0.0044$~\cite{Akrami:2018odb}.

In fact, the spectral index is sensitive to the detailed shape of the inflaton potential, and even a tiny correction 
can easily modify the predicted value of $n_s$ to give a better fit to data. To this end, we introduce
 a small but nonzero CP phase, $\theta$. 
For sufficiently small $\theta (>0)$,  one can again expand the potential around the origin
as 
\beq
 V_{\rm inf}(\phi) \simeq V_0 - \lambda \phi^4 - \Lambda^4 \theta \frac{\phi}{f_\phi} + \cdots.
\eeq
The linear term can effectively increase the predicted value of $n_s$,
because, if $\theta > 0$,  it shifts the inflaton field at the horizon exit of the pivot scale  to smaller values where the curvature of the potential, $|V''_{\rm inf}|$, is smaller~\cite{Takahashi:2013cxa}. Note that the linear term does not directly contribute to the $\eta$ parameter. 
A similar, but slightly weaker effect is obtained by varying the relative height of the potential, $\kappa$,  which induces a quadratic term. In fact, if the flatness of the inflaton potential is due to the anthropic selection of the parameters for the inflaton potential,
such nonzero values of $\theta$ and $\kappa-1$ may be reasonable because the successful slow-roll inflation only requires a sufficiently flat potential where the slow-roll parameters are smaller than unity.

\begin{figure}[!t]
  \begin{center}
   \includegraphics[width=105mm]{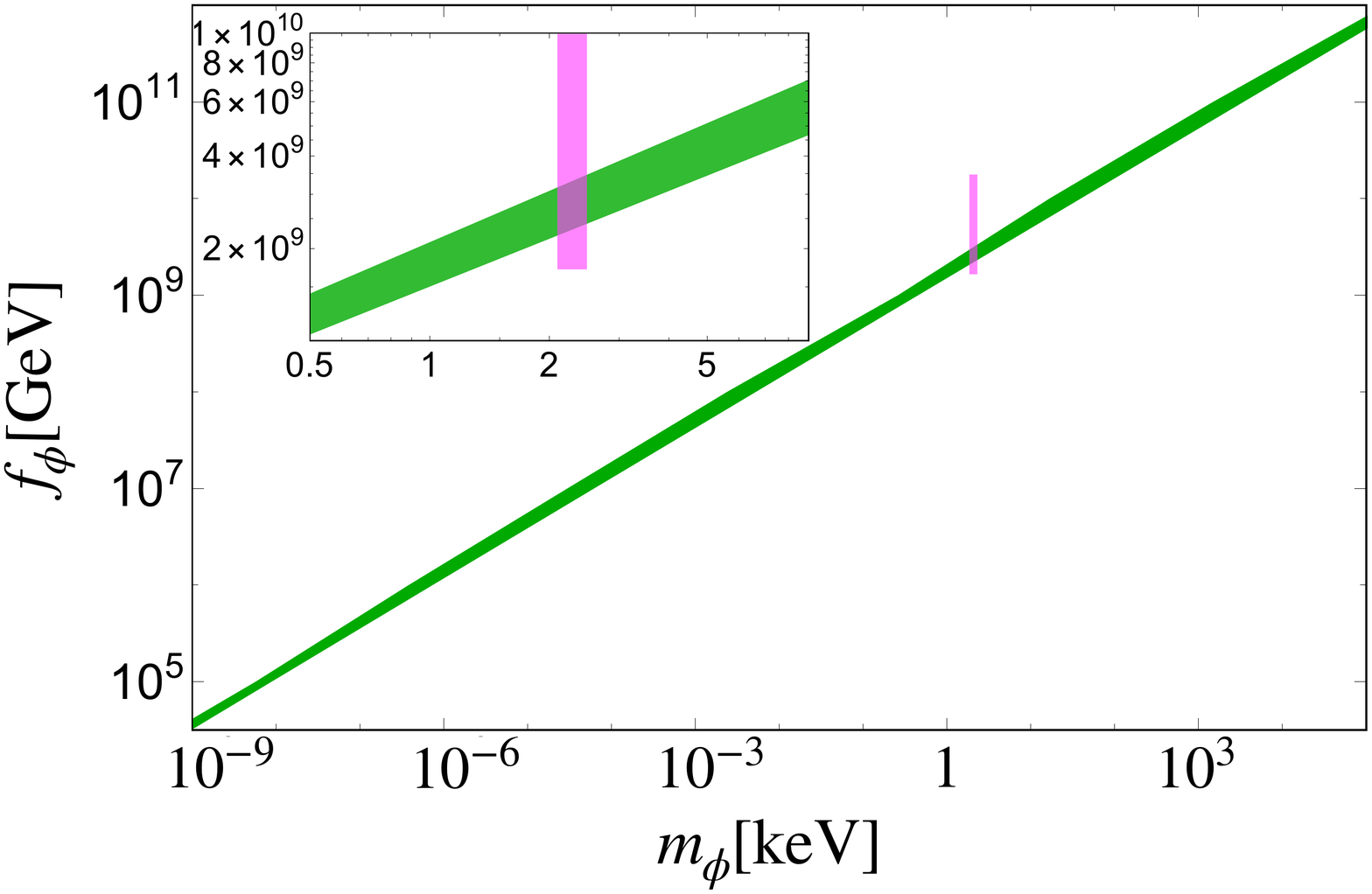}
\end{center}
\caption{
The predicted relation between $m_\phi$ and $f_\phi$ 
in the ALP inflation with $n=3$ is shown by the green line,
which is adapted from Ref.~\cite{Takahashi:2019qmh}. 
The mass and decay constant suggested by the XENON1T
excess are shown by the red shaded region for $q_e \sqrt{r}
=0.1  \, \text{-} \,  1$. As discussed in the text, the two regions overlap around $q_e \sqrt{r} \simeq 0.2$.
}
\label{fig:2}
\end{figure}

The inclusion of nonzero $\theta$ and $\kappa-1$ leads to an interesting
relation between the mass and the decay constant.
From Eq.\,\eqref{eq:nsform}, 
we obtain
\beq
H_{\rm inf} \approx \sqrt{\frac{V_0}{3 M_{p}^2}} \sim \sqrt{|V_{\rm inf}''(\phi_*)|} 
            \sim \sqrt{|V_{\rm inf}''(\phi_{\rm max})|},
\eeq 
where we have used in the last equality 
the fact that $\phi_*$ is close to the potential maximum located at $\phi = \phi_{\rm max}$ in the hilltop inflation.
Due to the upside-down symmetry, we have 
\beq
\label{mass=H}
m_{\phi} \sim H_{\rm inf},
\eeq 
where 
\beq
m_\phi \equiv \sqrt{V_{\rm inf}''(\phi_{\rm min})} 
\simeq  \left(\frac{9(n^2-1)}{2} \right)^\frac{1}{6} \theta^\frac{1}{3} \frac{\Lambda^2}{f_\phi}.
\eeq
From \eqref{pninni} and \eqref{mass=H}, we arrive at the relation~\cite{Czerny:2014xja,Daido:2017wwb, Takahashi:2019qmh,Daido:2017tbr}
\beq
f_\phi \sim 10^{3}\sqrt{m_\phi M_{p}}.
\eeq 
The relation was numerically confirmed in Refs.~\cite{Daido:2017wwb, Daido:2017tbr,Takahashi:2019qmh} by solving the inflaton dynamics taking account of higher order corrections to the spectral index.
We show in Fig.\,\ref{fig:2} the numerical result of \cite{Takahashi:2019qmh}.

For the current purpose we have fitted the numerical results 
 of  Ref.\,\cite{Takahashi:2019qmh}
for $n=3$ in the range of $10^{9}\GeV \leq f_\phi\leq 10^{11}\GeV$
and $|\k-1|\lesssim (f_\phi/M_{p})^2,|\theta|\lesssim(f_\phi/M_{p})^3$ to obtain
\begin{align}
\laq{fit1}
H_{\rm inf} &\simeq  (0.1 \, \text{-} \, 1)\KeV \sqrt{\frac{2-2/n^2}{16/9}}\left(\frac{f_\phi}{10^{9}\GeV}\right)^2 ,\\
\laq{fit2}
 \lambda &\simeq  (1 \, \text{-} \, 50)\times 10^{-13} \left(\frac{10^{9}\GeV}{f_\phi}\right)^{0.11},
\end{align}
where the $n$-dependence is reproduced analytically. 
Thus we arrive at
\begin{align}
\label{decay-mass}
f_\phi&\simeq (2 \, \text{-} \, 4)\times 10^9\GeV
\left(\frac{n}{3}\right)^\frac{1}{2}
\left(\frac{m_\phi}{2\,{\rm keV}}\right)^{\frac{1}{2}}.
\end{align}
 One can see that the mass (\ref{mass}) and decay constant (\ref{fphi})
suggested by the XENON1T excess satisfy the relation for $q_e \sqrt{r} \sim 0.1$.
 In the following we take $n = 3$, but our results are not
so sensitive to the choice of $n$, and the dependence on $n$
can be easily read from Eqs.\eqref{fit1}, \eqref{fit2} and \eqref{decay-mass}.

\begin{figure}[t!]
  \begin{center}
   \includegraphics[width=110mm]{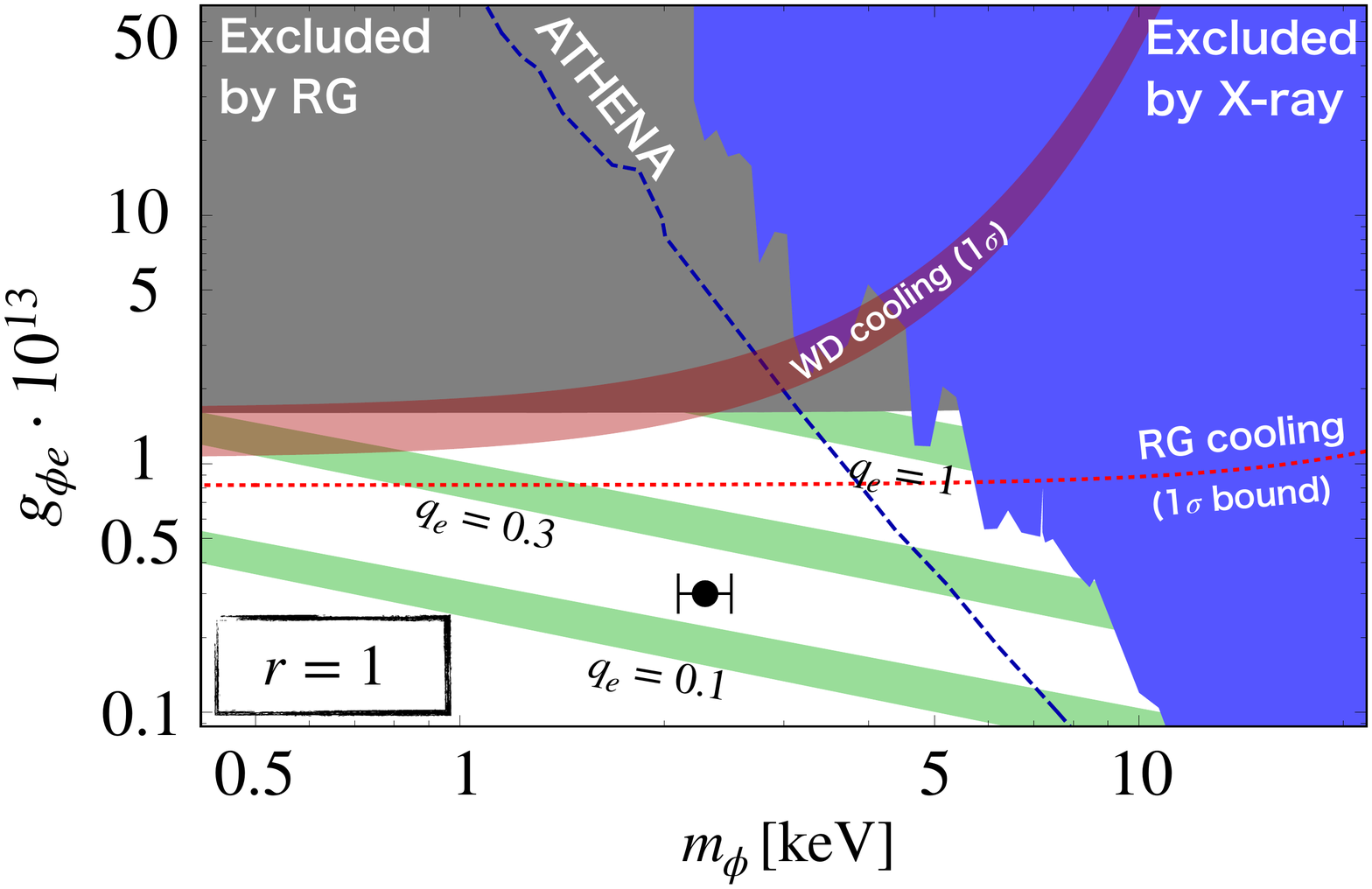}
   \includegraphics[width=110mm]{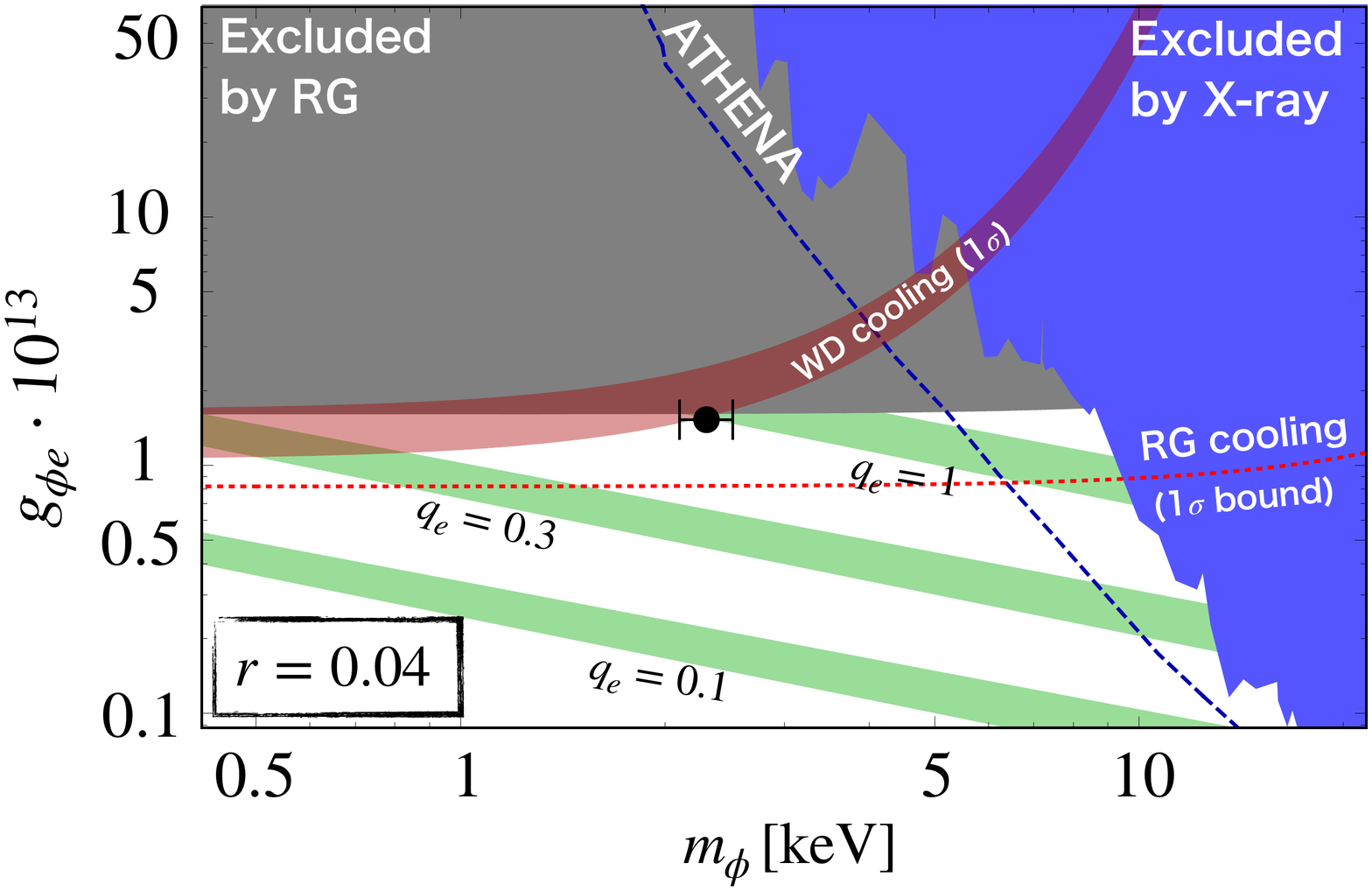}
\end{center}
\caption{
The mass and coupling to electrons suggested by the XENON1T excess
are shown by black dots for $r=1$ (upper panel) and $r=0.04$ (lower panel).
The predicted relation for $g_{\phi e}$ and $m_\phi$ in the ALP inflation is shown as green bands for 
 $q_e = 1, 0.3$, and $0.1$ from top to bottom.
 The X-ray bound for anomaly-free ALP DM~\cite{Nakayama:2014cza,Takahashi:2020bpq} is shown by the blue shaded
 region.
The future sensitivity of ATHENA is shown by the blue dashed line~\cite{Neronov:2015kca,Boyarsky:2018tvu}. 
The red shaded region is preferred  by the cooling of white dwarf stars. The gray region and the red dashed line represent the bound from the cooling of  the tip of the red-giant branch~\cite{Capozzi:2020cbu} at $2\s$ and $1\s$ level, respectively.
 }
\label{fig:3}
\end{figure}

In order to take a closer look, we show  in Fig.\,\ref{fig:3} 
the relation for different values of $q_e$ as
well as other cosmological and astrophysical constraints.
The upper and lower panels are for 
the fraction of ALP DM, $r = \Omega_{\phi}/\Omega_{\rm DM} =1$ and $0.04$,
respectively. The consistency relation (\ref{decay-mass}) are shown as green bands for the different PQ charge of electron, $q_e = 1, 0.3$ and $0.1$ from 
top to bottom. The mass  (\ref{mass}) and the electron coupling (\ref{gphie}) suggested by the XENON1T
excess are shown as black dots. We can see that, if the ALP DM responsible for  the XENON1T excess is the inflaton,
the PQ charge of electron and the fraction of ALP DM satisfy
\begin{align}
    q_e \sqrt{r} \simeq 0.2.
\end{align}
The blue shaded region is excluded by the X-ray observations, and the future sensitivity reach of ATHENA is shown by the blue dashed line~\cite{Neronov:2015kca,Boyarsky:2018tvu,Caputo:2019djj}.
Here we have adopted the decay rate into photons based on the
anomaly-free ALP DM model~\cite{Nakayama:2014cza,Takahashi:2020bpq}.
On the other hand,  the cooling argument
based on the tip of red-giant branch stars places a tight bound on $g_{\phi e}<1.60 \, (0.82) \times 10^{-13}$ at 95$\%$ (68$\%$) CL, and the  gray shaded region shows the $2\sigma$ excluded region, and the red dotted line is the 1$\sigma$ bound~\cite{Capozzi:2020cbu}.\footnote{Here we have simply adopted the Boltzmann suppression for the ALP emission from the stellar objects, assuming the typical
temperatures in white dwarf (red giant) is about $1 \KeV$ ($10\KeV$)~\cite{Calibbi:2020jvd}, but a dedicated analysis
might be necessary to derive the precise mass-dependence.}
The RG cooling bound gives a lower bound on the fraction of ALP DM as $r \gtrsim 0.01$. Note that the red shaded region is favored by the analysis of white dwarf luminosity function~\cite{Bertolami:2014wua},
which is close to the values suggested by the XENON1T excess especially
in the case of $r = 0.04$. In the rest of this section we discuss the reheating of the ALP inflation and the abundance of thermally produced ALPs.

\subsection{Reheating of the ALP inflation} 
\subsubsection{General argument}
After the slow-roll inflation, 
the ALP (inflaton) starts to oscillate around its potential minimum. 
The ALP mass at the minimum is of order keV, but it initially has a much larger
effective mass, and so,
it can efficiently transfer its energy to the SM particles.
Here we estimate how efficient the energy transfer should be for successful reheating. 

If the ALP couplings to the SM particles are sizable,
the reheating completes soon after the onset of oscillations.
In other words, the reheating is instantaneous. 
The reheating temperature in this case
is given by 
\begin{equation}
\label{TR}
T_{R} \simeq 
(3  \, \text{-} \,  8)
\times 10^{5}{\rm GeV}
\left( \frac{g_*(T_R)}{107.75} \right)^{-1/4}
\left(\frac{f_\phi}{10^{9}\GeV}\right),
\end{equation}
where we have used (\ref{fit1}) and set $g_*(T_R)=107.75$
counts all the SM degrees freedom plus thermalized ALP (inflaton)
(see later discussion for the thermal production of the ALP).
The detailed process of reheating, however, is quite non-trivial because the effective mass of the ALP decreases in time and thermal dissipation processes also play the important role. 

Let us approximate the ALP potential around the potential minimum as 
\begin{align}
V(\phi) &\simeq 
\frac12 m_\phi^2 (\phi - \phi_{\rm min})^2
+ \lambda(\phi - \phi_{\rm min})^4.
\end{align}
The quadratic term becomes relevant 
only after the oscillation amplitude of ALP becomes smaller than 
\beq
 \phi_c = \frac{m_\phi}{\sqrt{2\lambda}}. 
\eeq
The effective mass of the ALP is approximately given by 
\beq
 m_{\rm eff} \simeq 
\left\{
	\begin{array}{ll}
	\displaystyle{\sqrt{12 \lambda} \, \abs{\phi} }
	\qquad &{\rm for} \abs{\phi} > \phi_c \\
	&\\
	\displaystyle{m_\phi}
	\qquad &{\rm for} \abs{\phi} < \phi_c 
	\end{array}
	\right.,
\eeq
where $\abs{\phi}$ is the amplitude of the ALP oscillation. 
Soon after the onset of oscillations, the effective mass is about $3 \TeV$ for $\lambda = 10^{-12}$ and $\abs{\phi} = f_\phi = 10^{9} \GeV$,
and it can decay into relatively heavy SM particles. The decay products quickly thermalize and form
thermal plasma.
Because of the time-dependent effective mass, 
however, 
the perturbative decay stops at a certain point, and afterwards
the dominant reheating process is taken over by thermal dissipation process. 
If the dissipation process is efficient,  the ALP condensate will evaporate completely. If not, there may remain some amount of the ALP condensate which contributes to DM. In the former case, the ALP is considered to be thermalized, which we will discuss later in this section.

We denote the fraction of the remnant of the ALP condensate by $\xi$: 
\beq
\xi \equiv \left.\frac{\rho_\phi}{\rho_\phi + \rho_R}\right|_{\rm after\,reheating},
\eeq
where $\rho_\phi$ and $\rho_R$ are the energy density of the ALP condensate and radiation, respectively,
and $\xi$ is defined just after the reheating (decay or dissipation) becomes ineffective.
The precise value of $\xi$ can be estimated by numerically solving the Boltzmann equation 
for given interactions. Here let us estimate how small $\xi$ should be for successful reheating.

The amplitude of the ALP oscillations after the reheating is about $\xi^\frac{1}{4} f_\phi$ since
the initial amplitude is of order $f_\phi$.
Then, the ALP energy density to entropy ratio is given by
\begin{align}
\frac{\rho_\phi}{s} &\simeq \frac{3}{4} \Delta \xi^\frac{3}{4} \frac{m_\phi T_R}{\sqrt{2 \lambda} f_\phi},
\end{align}
where we have introduced an entropy dilution factor $\Delta (<1)$ for later purpose.
Using Eqs.\,\eqref{fit2} and \eqref{TR}, we obtain
\begin{align}
\label{mbelow}
\xi \sim  10^{-4} \times \left(\frac{10^{-3}}{\Delta}\right)^{4/3}  \left(\frac{2 \KeV}{m_\phi}\right)^{4/3}  \left(\frac{\Omega_{\rm \phi}h^2}{0.12}\right)^{4/3}
\left(\frac{f_\phi}{10^{9} \GeV}\right)^{-0.073}.
\end{align}
Since the ALP relic density should not exceed the observed DM density,
$\Omega_{\rm \phi}h^2 \lesssim 0.12$, we have an upper bound on $\xi$.
If there were not for any entropy dilution (i.e. $\Delta = 1$), $\xi$ should be
smaller than $10^{-8}$, and in the presence of the entropy dilution of $10^{-3}$,
it should be smaller than $10^{-4}$.

\subsubsection{Specific examples of the ALP couplings}

Now we specify  interactions between the ALP and SM particles to estimate $\xi$. 
As we consider the anomaly-free ALP model where the continuous shift symmetry (PQ symmetry) of $\phi$ is not broken by the interactions with the SM particles,  the relevant interactions  are given by the following derivative couplings with SM fermions\footnote{It is in principle possible to introduce the anomalous coupling of the ALP to
$Z,W^{\pm}$ gauge bosons as long as the anomalous coupling to photons is canceled. 
The induced photon coupling via the gauge boson loops should be suppressed by ${\cal O}(m_\phi^2/m^2_{W,Z})$ and the dominant one can be given by Eq.\,\eqref{photon}.
} 
\beq
\label{derivatives}
{\cal L}^{\rm int}_{\rm eff}= C_{ ij}\frac{\partial_\mu \phi}{2f_\phi} \bar{\psi}_i \gamma_5 \gamma^\mu {\psi}_j+{\cal O}(1/f_\phi^2),
\eeq
where $\psi_i$ represents the SM fermion with a flavor index $i$ running over
$i= e,\mu,\tau, u, d, s, \cdots$, and $C_{ ij}$ is a Hermitian matrix representing the ALP coupling.  Note that there are no anomalous couplings to
the SM gauge bosons by definition of the anomaly-free ALP. We have also concentrated on the CP-conserving interactions for simplicity.\footnote{By adding a small CP-violating interaction, $\tl{g}_{\phi e} \partial_\mu \phi \bar{\psi}_e \gamma^\mu {\psi}_e $, one can enhance the cooling rate of the WD and RG stars~\cite{Raffelt:1996wa}, 
although a pure CP-even scalar DM explanation is excluded by the RG cooling~\cite{Bloch:2020uzh}. Then one may be able to explain the
WD cooling without running afoul of the RG cooling constraint
due to the different core temperature even when the ALP explains all DM (see Fig.~\ref{fig:3}).
}

By performing field redefinition, we can rewrite the derivative couplings in terms of the Yukawa couplings,
\begin{equation}
\label{int1}
{\cal L}_{\rm Yukawa}=\frac{ig_{\phi {ij}}}{v} \phi \( \bar{\p_{i}}\gamma_5 H \p_j \) 
\end{equation}
with
$g_{\phi ij }= (y^\dagger_\psi\cdot C+C \cdot 
y_\psi)_{ij}\,v/(2f_\phi)$, where $y_\psi$ represents the Yukawa matrices, 
$v\approx 174\GEV$ is the Higgs vacuum expectation value (VEV), and $H$ should be replaced with $\tilde{H} = i \sigma_2 H^*$ for up-type quarks. We also dropped the term of ${\cal O}((\phi/f_\phi)^2)$. The electron coupling is given by
 $g_{\phi e}\equiv g_{\phi ee}$ in this notation.  
In this basis, anomalous couplings to the SM gauge bosons may or may not be induced by the change of path-integral measure when one performs the field redefinition
from (\ref{derivatives}) to (\ref{int1}).
Whether an anomalous coupling appears depends on the nature of the PQ current in (\ref{derivatives}), and therefore on the UV origin of the anomaly-free PQ symmetry. For instance,  if the PQ anomaly is cancelled among the contributions of the SM fermions, in other words, the PQ current in (\ref{derivatives}) is anomaly-free, 
the anomalous coupling does not appear  above the EW scale in the basis of (\ref{int1}).
This is the case of the lepton specific \cite{Nakayama:2014cza,Takahashi:2020bpq} or the two-Higgs doublet \cite{Takahashi:2019qmh,Takahashi:2020bpq} model. 
On the other hand, if the anomaly is cancelled between the SM fermions and 
a PQ fermion which may be as heavy as the PQ scale, there appear anomalous couplings below the PQ scale in (\ref{int1}). The couplings are originated from the triangle diagram contributions of integrating out the PQ fermion. 
When we further integrate out all the SM fermions to see the physics at around keV scales, the anomalous couplings will be cancelled out.

With any UV completion, we can move to the basis where ALP couples to fermions only via derivative couplings as \eqref{derivatives}. In this basis, there should not be any additional anomalous couplings to photons or gluons since the ALP is anomaly-free to them. 
By integrating out the fermions in the effective Lagrangian \eqref{derivatives},  the ALP-photon coupling is generated accompanied with the derivatives of $\phi$, e.g. $\partial^2\phi F\tilde{F}$, thanks to the shift symmetry. Since the electron is the lightest charged fermion, the dominant photon coupling is \eq{photon} unless $q_e$ is highly suppressed compared with the PQ charge for the other fermions~\cite{Nakayama:2014cza,Takahashi:2020bpq} (see also Ref.~\cite{Pospelov:2008jk}).

Now let us discuss reheating in this set-up. In the Yukawa basis we generically have two sources for the reheating, the couplings to fermions and to gauge bosons. 
In fact, even if the anomalous couplings are present in the Yukawa basis, 
the reheating process via the anomalous couplings turns out to be negligible in the parameter space
under consideration~\cite{Daido:2017wwb, Daido:2017tbr}.
Thus, in the following we discuss the reheating process through \eq{int1}.
To simplify the discussion, we consider flavor-diagonal couplings
\beq
C_{ij}=q_{i} \delta_{ij},~~ y_{\psi}=\delta_{ij} y_{\psi_i},
\eeq 
and we neglect the off-diagonal components of the Yukawa matrices.

The ALP decays into two fermions, 
$\phi\rightarrow \psi_i + \bar\psi_i$, with the decay rate of 
\begin{equation}
\laq{decf}
\G_{{\rm dec},\psi}\simeq
\frac{n_c }{8\pi } \(\frac{q_{i} m_{\psi_i} }{f_\phi}\)^2m_{\rm eff},
\end{equation}
if the mass or thermal mass of $\psi$ is smaller than the effective mass of the ALP. 
On the other hand, if the thermal mass of $\psi$ exceeds $m_{\rm eff}$, namely $eT\gtrsim m_{\rm eff}$ (or $g_s T\gtrsim m_{\rm eff}$), 
the decay process stops and the dissipation process
such as 
$\phi+ \psi_i \rightarrow \psi_i + \gamma \,({\rm or~} g)$ becomes relevant. The dissipation rate is given by~\cite{Mukaida:2012bz} 
\begin{equation}
\label{disbrF}
\G_{{\rm dis},\psi}\simeq \displaystyle{ A_0 n_c\(\frac{q_{i} m_{\psi_i}  }{f_\phi}\)^2 \(\frac{ \a_\psi}{2\pi^2}\) T,}
\end{equation} 
for $T < T_{\rm EW}$, where $T_{\rm EW}$ is the temperature of the electroweak phase transition,
$A_0$ ($\simeq 0.5$) is a numerical constant, $\alpha_\psi$ is the fine-structure constant of the relevant gauge coupling, $e$ (or $g_s$). 
Before the electroweak phase transition, on the other hand, 
the dissipation proceeds via the interaction with the Higgs field such as 
$\phi+{\rm Higgs} \rightarrow \psi_i + \bar \psi_i$. 
Then the dissipation rate is given by~\cite{Daido:2017tbr,Salvio:2013iaa}
\begin{equation}
\label{eq:disF}
\G_{{\rm dis},H}\simeq  
n_c  \(\frac{q_i^2 y_{\psi_i}^2  }{2 \pi^3 f_{\phi}^2}\) T^3, 
\end{equation} 
for $T > T_{\rm EW}$. 
Note that, while the perturbative decay gradually becomes inefficient as
the effective mass decreases in time, 
the dissipation rate of (\ref{eq:disF}) and (\ref{disbrF}) is 
independent of $m_{\rm eff}$, and so, it is efficient until the ALP disappears completely.

Now we can calculate the fraction of the relic ALP condensate 
$\xi$ by solving 
the Boltzmann equations of
\begin{align}
\left\{
	\begin{array}{ll}
	\displaystyle{\dot{\rho}_\phi+4H\rho_\phi=-\Gamma_{\rm tot}\rho_\phi} \\
	&\\
	\displaystyle{\dot{\rho}_r+4H\rho_r=\Gamma_{\rm tot}\rho_\phi}
	\end{array}
	\right.,
\end{align}
where $\Gamma_{\rm tot} = \G_{{\rm dec},\psi}+\G_{{\rm dis},\psi (H)}$. 
Since $\G_{{\rm dis},\psi H} / H$ is larger for higher $T$, we find that the dissipation effect is most efficient just after 
the inflation ends.
This implies that the remnant fraction of the ALP condensate 
is exponentially suppressed as 
\beq
 \xi \propto \exp \lkk - \G_{{\rm dis}, H} / H \rkk
\eeq
with
\beq
 \frac{\G_{{\rm dis}, H} }{ H} 
 &\simeq& 
 n_c  \(\frac{q_i^2 y_{\psi_i}^2  }{2 \pi^3}\) 
 \sqrt{\frac{90}{\pi^2 g_*(T_R)}}
 \frac{T_R \Mpl}{f_\phi^2} 
 \\
 &\simeq& 10^4 \times n_c q_i^2 y_{\psi_i}^2 
 \left( \frac{T_R}{10^6 \GeV} \right) 
 \left( \frac{f_\phi}{10^9 \GeV} \right)^{-2}
 \left( \frac{g_*(T_R)}{107.75} \right)^{-1/2}. 
 \label{reheating}
\eeq
Thus the remnant of the ALP inflaton is negligibly small 
if $n_c q_i^2 y_{\psi_i}^2 \gtrsim 10^{-3}$. 
For $q_i={\cal O}(0.1 \, \text{-} \, 10)$, $\xi$ becomes sufficiently small and the reheating will be successful  if the $i$ contains (relatively) heavy fermions 
such as $\tau, b, c$ or $t$.

We have numerically solved the Boltzmann equation and find that
the reheating becomes inefficient and $\xi \approx 1$ for $n_c q_i^2 y_{\psi_i}^2 \ll 10^{-3}$ and $f_\phi \sim 10^9\,$GeV, even if we take into account the dissipation effect of \eq{disbrF} and the perturbative decays at a later time. 
For $n_c q_i^2 y_{\psi_i}^2 \gtrsim 10^{-3}$, we have confirmed that
$\xi$ is exponentially suppressed and the reheating completes instantaneously. The reheating temperature is therefore given by
\eq{TR}. 

In summary, if the XENON1T excess is explained by the ALP DM which drives
inflation in the early universe, the ALP must be coupled to heavy fermions for successful reheating. The reheating is almost instantaneous, and the reheating temperature is of order $10^6$\,GeV.

\subsection{Thermal production of ALP}

The ALP can also be produced from thermal plasma through the inverse process of the diffusion effect.
Since the diffusion effect is efficient for the successful reheating, the ALP is always thermally populated. This is a rather robust prediction of our scenario. The thermally produced ALP can be the potential source of the XENON1T excess.

The abundance of the thermally produced ALP is given by 
\beq
\label{Omega_th}
 \Omega_\phi^{\rm (th)} h^2 \simeq 2.1 \Delta \lmk \frac{m_\phi}{2 \KeV} \rmk \lmk \frac{g_{*s}}{106.75} \rmk, 
\eeq
where $g_{*s}$ is the relativistic degrees of freedom for entropy density of the SM particles and we have included the entropy dilution factor $\Delta (<1)$.
In order not to exceed the observed DM abundance for $\Delta = 1$, 
the thermally produced ALP must be diluted by some late-time 
entropy production.
The required amount of the entropy dilution factor is given by
\beq
 \Delta \simeq 0.06 \lmk \frac{\Omega_\phi^{\rm (th)} h^2}{0.12} \rmk, 
 \label{delta}
\eeq
where we used $m_\phi \simeq 2 \KeV$. In the next section we will discuss such entropy production mechanism.

Lastly, we note that a thermally produced ALP with a keV-scale mass has a nonzero free-streaming velocity at the structure formation, and it behaves as warm DM.
The Lyman-$\alpha$ observation puts an upper bound on the free-streaming velocity, which can be translated to the
upper bound on the mass of the warm DM for a given momentum distribution. In the case of the sterile neutrino, its mass is
bounded as $m_{\nu_s}> 5.3$ keV~\cite{Viel:2005qj,Irsic:2017ixq}
if it constitutes the dominant DM. 
The bound on the ALP warm DM should be of the same order,
although the degree of freedom and the statistics are different.
Also, there are constraints from observations of CMB, the baryon acoustic oscillation (BAO) and the number of dwarf satellite galaxies in the Milky Way~\cite{Diamanti:2017xfo}, which imply that the ALP warm DM with mass of $2$ keV is likely in a tension with observations. One way to relax the tension is to consider a case that
the ALP occupies only a fraction of DM. Indeed, if $r < 0.1$,
the constraints are significantly relaxed.
To explain the XENON1T excess, we need $r > 0.01$ (see Fig.\,\ref{fig:3}). Therefore, our scenario may be tested if the aforementioned bounds for the warm DM are  improved~\cite{Sitwell:2013fpa, Kogut:2011xw, Abazajian:2016yjj,Baumann:2017lmt}.

If the thermally produced ALP constitutes only a fraction of DM, the dominant component may come from the remnant of the inflaton condensate via incomplete reheating (see  \eqref{mbelow}).
This is the case if $n_c q_i^2 y_{\psi_i}^2 \sim 10^{-3}$. 
On the other hand, the dominant DM component may be another axion(s).
As we will discuss later, there may be many axions in nature as in the axiverse scenario, and then, 
$m_\f \sim H_{\rm inf}\sim$\,keV may be explained by the anthropic argument based on the initial axion amplitude and the abundance.

\section{Entropy dilution and leptogenesis}
\label{sec:entropy}

It is natural to introduce right-handed neutrinos 
to explain the smallness of the active neutrino masses by the seesaw mechanism and to produce baryon asymmetry via leptogenesis. 
The Lagrangian is written as 
\beq
\label{nuint}
 {\cal L}_{N_R} = ({\rm kinetic terms}) - \frac12 M_i 
 \bar{N^c_i} N_i - y_{ij}^{\nu} \bar{L}_j H \hat{P}_R N_i + {\rm h.c.},
\eeq
where $L_j$ is a left-handed Lepton field in the chiral representation.
As we will see shortly, the decay of right-handed neutrinos can result in an entropy dilution that is required in our scenario.

The right-handed neutrino, $N_i$ is also produced through the inflaton decay if it is lighter than $m_{\rm eff}$ and has derivative couplings like
\beq
{\cal L}^{\rm int}_{\rm eff}\supset  C_{N_iN_j} \frac{\partial_\mu \phi}{2f_\phi} \bar{N_i}\gamma_5\gamma^\mu {N_j}. 
\eeq
In the mass basis, we obtain, 
\beq
{\cal L}_{\rm mass}= i C_{N_{i} N_j}(M_{i}+M_{j})\frac{\phi}{2f_\phi} \bar{N^c_i} \gamma_5 N_j.
\eeq
We have ignored the ALP coupling 
via Yukawa interactions in Eq.~\eqref{nuint} since 
$y^{\nu}_{ij}$ is small in the parameter region of our interest. 

In particular, the right-handed neutrino, $N_1$, corresponding to the lightest active neutrino, can be produced. 
The number density of $N_1$ to entropy density ratio is given as
\beq
\label{nNs}
\frac{n_{N}}{s}\sim \frac{3}{4} \frac{T_R}{m_{\rm eff}/2} \frac{\G_{\phi\to N_1 N_1}}{\G_{\rm tot}}\sim 7 \frac{C_{N_1N_1}^2 M_{1}^2}{n_c(q_i y_{\psi_i})^2 T_R^2}. \eeq
This takes the maximum value of ${\cal O}$($10^{-3}$), when $M_1 \sim m_{\rm eff}$ and $n_c (y_{\psi_i} q_i)^2\sim 10^{-3}$.\footnote{Note that, toward the end of the reheating, the amplitude of inflaton decreases and the effective mass of the inflaton may become smaller than $M_1$. Then the inflaton decay to $N_1$ is kinematically forbidden. However this does not
modify the abundance of $N_1$ significantly from (\ref{nNs}).}
The right-handed neutrino comes to dominate the Universe when the temperature decreases to 
\beq
 T_{\rm dom} \simeq \frac{4 }{3}M_1 \frac{n_N}{s}  \sim 10\GeV\left(\frac{10^{-3}}{n_c (y_{\psi_i} q_i)^2/C_{N_1N_1}^2 }\right)\left(\frac{10^6\GeV}{T_R} \right)^2 \left(\frac{M_1}{10^3\GeV}\right)^3. 
\eeq

Then 
$N_1$ decays into the SM particles to reheat the universe with a temperature $T=T_N$.\footnote{For successful entropy production,
the decays of $N_1\to N_{2,3}+\phi$ should be suppressed. This is either kinetically forbidden with the condition of $M_{2,3}\gtrsim M_1$ or suppressed by the small off-diagonal component of $C_{N_i N_j}$.} 
The  ALP-number-to-entropy ratio after the decay of $N_1$ is given as 
\beq
\left.\frac{n_\phi}{s}\right|_{T=T_N}=
\left.\frac{n_N}{s}\right|_{ T=T_{N}}
\left.\frac{n_\phi}{n_N}\right|_{ T=T_{\rm dom}}  = \frac{T_N}{T_{\rm dom}}  \left.\frac{n_\phi}{s}\right|_{ T=T_{\rm dom}}. 
\eeq
Thus, we identify the dilution factor as $\Delta=T_N/T_{\rm dom}.$ 
The right-handed neutrino should decay before the big bang nucleosynthesis (BBN), $T_N\gtrsim 1\MeV$,
not to spoil the successful prediction of the light element abundances~\cite{Kawasaki:1999na,
Kawasaki:2000en, 
Hannestad:2004px, Ichikawa:2006vm, DeBernardis:2008zz, deSalas:2015glj, Hasegawa:2019jsa}. Then, the dilution factor is bounded below
\beq 
 10^{-4}  \lmk \frac{10 \GeV}{T_{\rm dom}}\rmk \lesssim  \Delta \leq  1,
\eeq
which in turn implies that the abundance of the thermally produced ALP is in the range of
\beq
\Omega_\phi^{\rm (th)}= O(10^{-3} \, \text{-} \, 1).
\eeq 
This has a large overlap with the range of $r$ where the XENON1T excess can be explained
(see Fig.\,\ref{fig:3}).

Let us mention a possible  baryogenesis associated with the right-handed neutrinos.
In the case of $ 1\,{\rm TeV} \lesssim M_2 \approx M_3 \lesssim 10^6\GeV$,
the right amount of baryon asymmetry can be generated via
resonant leptogenesis~\cite{Pilaftsis:1997dr,Buchmuller:1997yu},
though thermal leptogenesis~\cite{Fukugita:1986hr} is difficult because of the low reheating temperature $T_R\sim 10^6\GeV$. 
On the other hand, in the case of $M_{2,3}\sim 1 \, \text{-} \, 100\GEV$, the baryon asymmetry can be obtained from the flavor oscillation of the left-handed leptons which are produced from the inflaton decay if the ALP-lepton derivative couplings are flavor-violating and are not too small~\cite{Hamada:2018epb,Eijima:2019hey}.\footnote{In general,  the inflaton-lepton couplings involves lepton flavor violation, which
may be searched for by  the future measurement of $\mu\to e+\phi$~\cite{Calibbi:2020jvd,1807738}.}
Note that generated baryon asymmetry is also diluted by the entropy production
of $N_1$, and one needs to generate larger baryon asymmetry by a factor of $\Delta^{-1}$ than
the observed value.

Another way to have  late-time entropy production is to introduce a B$-$L Higgs field. The U(1)$_{\rm B-L}$ gauge symmetry is assumed to be restored after inflation, and the B$-$L Higgs field stays at the origin and dominates the universe for a while.
After the symmetry breaking, the B$-$L Higgs decays into the SM particles and produce entropy that leads to $\Delta\sim {\cal O}(0.01 \, \text{-} \, 0.1)$. See Appendix for the estimation of the dilution factor in this scenario.  In this case the right-handed neutrino can be produced efficiently in the symmetric phase through the U(1)$_{\rm B-L}$ gauge interactions.  The  decays of the right-handed neutrinos can generate baryon asymmetry via resonant leptogenesis.

\section{Discussion} 
\label{sec:discussion}

So far we have considered a single-field inflation model, but the ALP inflation can be extended to multi-field inflation where the two axions respectively
play the role of the inflaton and waterfall field as in the hybrid inflation~\cite{Daido:2017wwb}
(see also Refs.~\cite{Peloso:2015dsa,Carta:2020oci}).
Even in this case, the consistency relation between the mass and decay constant 
is known to hold, while the reheating process can be quite different~\cite{Daido:2017wwb}. 
It may be able to induce the entropy production from one axion while the other becomes DM and explains the XENON1T excess.

In our scenario, 
the inflation scale is predicted to be around $H_{\rm inf}\sim 1\,$keV. Since this is lower than the QCD scale, if the QCD axion, $a$, exists to solve the strong CP problem, it acquires a potential via the non-perturbative effect during inflation.
In the ALP inflation model considered in the text, the eternal inflation can take place around the potential maximum. 
If the inflation lasted sufficiently long, then the QCD axion field follows an equilibrium probability distribution peaked around the CP-conserving minimum with a variance $\sim 0.1 H_{\rm inf}^2(m_{a} f_{a})^{-1}$ 
due to the stochastic process, where $m_a \AND f_a$ are the QCD axion mass and decay constant, respectively.  The resulting abundance of the QCD axion is highly suppressed~\cite{Graham:2018jyp, Guth:2018hsa}.\footnote{If there is a mixing between the ALP and other axions, on the other hand, this is not necessarily the case because the minimum could be shifted after inflation~\cite{Guth:2018hsa}. See also Refs.~\cite{Daido:2017wwb,Co:2018mho,Takahashi:2019pqf, Nakagawa:2020eeg,Huang:2020etx}.}

Here let us comment on a possibility that the ALP mass is determined by the anthropic principle~\cite{Davies, Barrow, Barrow:1988yia, Hawking:1987en, Weinberg:1987dv}. 
In the string axiverse, there are many ALPs with the decay constants of order $10^{16}\GeV$~\cite{Green:1987sp,Svrcek:2006yi}. 
We expect that the masses of these ALPs are logarithmically distributed over a wide range of energy scales,
because their masses are generated by non-perturbative effects. 
Then, while heavy ALPs already settle down at their potential minimum during inflation, 
light ALPs may start to oscillate coherently around their potential minimum after inflation.
The energy densities of the coherent oscillations behave like DM. 

Now we shall take into account the anthropic condition on the DM abundance. 
If the DM is too abundant, 
the cosmological structure would be significantly complicated,
and the number of small-scale structure, like galaxies, might become quite large. 
As a result, the collision rate of stars/comets may be too large and there may be no time 
for sapient life to emerge on a planet like the Earth. 
The consideration along this line puts an upper bound on the DM abundance, which turns out to be
 larger than the observed DM abundance, but  not by many orders of magnitude. (see, e.g., Ref.~\cite{Tegmark:2004qd}). 
Then, the energy densities of the ALP coherent oscillations  must be suppressed to satisfy
the anthropic condition on the DM abundance. 

One might think that the initial misalignment angle of each ALP can be fine-tuned 
to satisfy the anthropic condition. 
However, the total amount of fine-tuning would be very large if there are many ALPs,
and it depends on which parameters are actually environmental parameters whether such tuning is justified.
Instead, we consider the case in which the anthropic condition is satisfied by varying the Hubble parameter during inflation, namely $H_{\rm inf}$. 
In Ref.~\cite{Ho:2019ayl}, Ho and two of the present authors (FT and WY) 
studied the stochastic behavior of string axions and showed that the abundance of the ALP
coherent oscillations is sufficiently suppressed if $H_{\rm inf} \lesssim {\cal O}(1) \KeV$ for
 the ALP decay constants of order $10^{16} \GeV$. 
Therefore, if $H_{\rm inf}$ is randomly distributed in the string landscape 
and is biased toward a larger value, 
$H_{\rm inf} = {\cal O}(1) \KeV$ is chosen by the anthropic condition on the DM abundance.  
Noting that the mass of the inflaton ALP is of the same order with $H_{\rm inf}$ in our scenario, 
this is another interesting coincidence with the XENON1T excess.

Lastly let us comment on a possibility of the PQ symmetry restoration. In the main text we have focused on the
case that the PQ symmetry is already broken during inflation as the inflaton was identified with the corresponding
ALP. In a more general case where the ALP is not necessarily the inflaton, the PQ symmetry could be restored 
in the early universe and gets spontaneously broken some time after inflation. This scenario has  advantages;
the ALP has no isocurvature perturbation, and the inflation scale can be high; also, the reheating temperature (of the SM sector) can be high which makes it easier to generate the right amount of baryon asymmetry. The cosmic strings and domain walls are formed after the phase transition, but they will disappear if the corresponding domain wall number is equal to unity, namely if there is a unique vacuum.
On the other hand, we need to make sure that the ALP is not thermalized, since its abundance would be larger
than the observed DM abundance and some entropy dilution would be required. In particular, if the PQ sector including the PQ scalar(s) are thermalized, the ALP is expected to be already thermally populated right after the phase transition. In the simplest case with a single PQ complex scalar, the thermalized  real and imaginary components will become 
the radial direction of the PQ-symmetry breaking field
and ALP after the phase transition. One way to avoid the overproduction of the ALP is to assume that
the PQ sector has been always decoupled from the SM sector, and it is never thermalized. Then, the timing of the PQ phase transition is model-dependent, and it depends on e.g. the detailed shape of the potential for the PQ scalar, or the interactions among the PQ scalars. If the phase transition takes place at a late time, the temperature of the SM sector can be so low that thermal production of the ALPs becomes negligibly small. In an extreme case in which the PQ sector has a clockwork structure, the phase transition indeed takes place at a very late time, and the resultant topological defects have a complicated structure~\cite{Higaki:2015jag,Higaki:2016yqk,Higaki:2016jjh}. Note, however, that 
the radial direction of the PQ-symmetry breaking field 
should dominantly decay to the SM particles so that hot/warm ALP is not overproduced~\cite{Jeong:2012np}.

\section{Conclusions}
\label{sec:conclusion}

In this paper we have pointed out that, when the XENON1T excess is interpreted in terms of
the ALP dark matter, the suggested mass and decay constant agree very well with the 
relation of \eqref{ALPinfrel} that is predicted in the ALP inflation model.
The relation was derived solely based on the CMB normalization and the scalar spectral index.
For the mass and decay constant suggested by the XENON1T excess, 
the Hubble parameter during inflation is predicted to be $H_{\rm inf}\sim \KeV.$
If the ALP is coupled to relatively heavy SM fermions, 
the reheating proceeds almost instantaneously 
and the reheating temperature  is given by
$T_R\sim 10^6\GeV$.

After the reheating, the ALP is  produced from the scattering in the thermal plasma. 
The thermally produced ALP DM explains the XENON1T excess 
once its abundance is diluted by a factor of $10$. Such a mild entropy dilution can be realized in simple cosmological scenarios. We have presented a scenario where
 the inflaton decays into right-handed neutrinos, and the lightest right-handed neutrino
 can  dominate and reheat the Universe to produce the entropy. 
Our scenario have various implications for small-scale structure as well as lepton flavor violation. 
Future observations of 21 cm lines, CMB, BAO, dwarf galaxies and $\mu\to e +$ missing will
provide us with further information on the possible relation between DM and inflation.

%
\section*{Acknowledgments}
F.~T. was supported by JSPS KAKENHI Grant Numbers
17H02878, 20H01894 and by World Premier International Research Center Initiative (WPI Initiative), MEXT, Japan. 
M.~Y. was supported by Leading Initiative for Excellent Young Researchers, MEXT, Japan. W.~Y was supported by JSPS KAKENHI Grant No. 16H06490.
%

\appendix

\section{Entropy dilution by thermal inflation}

Here we discuss another possible source of entropy dilution. 
Suppose that there is U(1)$_{\rm B-L}$ gauge symmetry, 
which is spontaneously broken at an intermediate scale to give a nonzero mass for the right-handed neutrino. 
We assume that the U(1)$_{\rm B-L}$ breaking field $\Psi$ has the Higgs-like potential such as 
\beq
 V(|\Psi|) = \lambda_{\Psi} \lmk \abs{\Psi}^2 - \frac12 v_\Psi^2 \rmk^2 + V_T(\Psi) \,. 
\eeq
Here, we include the thermal potential $V_T$, which is given by 
\beq
 V_T (|\Psi|) = \lmk \frac{\lambda_{\Psi}}{3} + q_\Psi^2 \frac{g^2}{4} \rmk T^2 \abs{\Psi}^2 \,, 
\eeq
where $q_\Psi$ ($g$) is the U(1)$_{\rm B-L}$ charge (gauge coupling) of $\Psi$. 
The critical temperature, below which the  U(1)$_{\rm B-L}$ Higgs field has a nonzero VEV, is then  given by 
\beq
 T_c = \sqrt{\frac{\lambda_{\Psi} }{\lambda_{\Psi}/3 + q_\Psi^2 g^2/4}} \, v_\Psi \,. 
\eeq

The entropy is released from the decay of the U(1)$_{\rm B-L}$ Higgs field. 
We define a dilution factor as the ratio of the final to the initial comoving 
entropy density as 
\beq
 \Delta^{-1} \equiv \frac{s_i a_i^3}{s_f a_f^3} = 
 1 + \frac43 \frac{g_{* s} (T_{\rm RH, \Psi})}{g_{*} (T_{\rm RH, \Psi}) \, T_{\rm RH, \Psi}} \frac{V (0)}{(2 \pi^2/45) \, g_{* s} (T_c) \, T_c^3 } \,, 
\eeq
where 
$g_{* s}$ ($g_{*}$) is the effective number of relativistic degrees of freedom for entropy (energy) density as a function of $T$. 
Here, 
$V(0) = \lambda_{\Psi} v_\Psi^4 / 4$, and 
$a_i$ ($a_f$) is the scale factor and 
$s_i$ ($s_f$) is the entropy density before (after) the thermal inflation. 
If the reheating completes instantaneously, the reheating temperature is given by 
\beq
T_{\rm RH, \Psi} = \lmk \frac{30 V(0)}{g_{*} (T_{\rm RH, \Psi}) \, \pi^2} \rmk^{1/4}. 
\eeq
Then, we obtain
\beq 
\Delta \simeq 0.06 \times (q_\Psi g)^{-3} \lmk \frac{g_{*s}(T_c) \lambda_{\Psi}}{10^{-3}} \rmk^{3/4},
\eeq  
where we have assumed $g_{* s} (T_{\rm RH, \Psi})= g_{*}(T_{\rm RH, \Psi})$, 
$\Delta \ll 1$, and $\lambda_{\Psi} \ll q_{\Psi}^2 g^2$.
For instance, if we take a value of $\lambda_\Psi \lesssim g^4 q_\Psi^4/16\pi^2$ which is slightly smaller than the natural value determined by the radiative correction, we obtain $\Delta \lesssim {\cal O}(1)$. Therefore, with a mild tuning of the $\lambda$ with respect to the Coleman-Weinberg set-up, $\Delta$ can be of order $0.01-0.1$.

\vspace{1cm}

\bibliography{references}

\end{document}